\documentclass{emulateapj}

\usepackage{natbib}
\usepackage{comment}
\usepackage{url}
\usepackage{color}

\newcommand{\etal}{et al.}
\newcommand{\mic}{\mathrm{\mu s}}

\shorttitle{Correlation of \textit{Fermi} photons with high-frequency radio GPs from the Crab pulsar}
\shortauthors{A.~V.~Bilous \etal}

\begin{document}

\title{Correlation of \textit{Fermi} photons with high-frequency radio giant pulses from the Crab pulsar}

\author{A.~V.~Bilous\altaffilmark{1},
V.~I.~Kondratiev\altaffilmark{2,3},
M.~A.~McLaughlin\altaffilmark{4,7},
S.~M.~Ransom\altaffilmark{5},
M.~Lyutikov\altaffilmark{6},
M.~Mickaliger\altaffilmark{4},
G.~I.~Langston\altaffilmark{7}}

\altaffiltext{1}{Department of Astronomy, University of Virginia, PO Box 400325, Charlottesville, VA 22904; avb3k@virginia.edu}
\altaffiltext{2}{Netherlands Institute for Radio Astronomy (ASTRON), Postbus 2, 7990 AA Dwingeloo, The Netherlands; kondratiev@astron.nl}
\altaffiltext{3}{Astro Space Center of the Lebedev Physical Institute, Profsoyuznaya str. 84/32, Moscow 117997, Russia}
\altaffiltext{4}{Department of Physics, West Virginia University, Morgantown, WV 26506}
\altaffiltext{5}{National Radio Astronomy Observatory, Charlottesville, VA 22903}
\altaffiltext{6}{Department of Physics, Purdue University, West Lafayette, IN 47907}
\altaffiltext{7}{National Radio Astronomy Observatory, Green Bank, WV 24944}

\begin{abstract}
  To constrain the giant pulse (GP) emission mechanism and test the
  model of Lyutikov (2007) for GP emission, we have carried out a
  campaign of simultaneous observations of the Crab pulsar at
  $\gamma$-ray (\textit{Fermi}) and radio (Green Bank Telescope) wavelengths.
  Over 10 hours of simultaneous observations we obtained a sample of
  2.1$\times 10^4$ giant pulses, observed at a radio frequency of 9\,GHz, and 77 \textit{Fermi}
  photons, with energies between 100\,MeV and 5\,GeV. The majority of GPs came from the interpulse (IP)
  phase window. We found no change in the GP generation rate within 10$-$120\,s windows at lags
  of up to $\pm 40$\,min of observed $\gamma$-ray photons. The 95\% upper
  limit for a $\gamma$-ray flux enhancement in pulsed emission phase window around all GPs is 4
  times the average pulsed $\gamma$-ray flux from the Crab.  For the subset of IP
  GPs, the enhancement upper limit, within the IP emission window, is
  12 times the average pulsed $\gamma$-ray flux. These results suggest that GPs, at least
  high-frequency IP GPs, are due to changes in coherence of radio emission 
  rather than an overall increase in the magnetospheric particle density.

\end{abstract}


\section{Introduction}

The Crab pulsar was discovered by \citeauthor{staelin1968} in
\citeyear{staelin1968} by its remarkably bright giant pulses (GPs).
Giant pulses are short (from few ns to few $\mic$), sporadic bursts of
pulsar radio emission~\citep{popov_stappers2007, hankins2003}. The nature of GPs is far from being clear and even the precise definition of giant pulse had not yet been given \citep{knight2006}. GPs generally occur only in certain narrow ranges of pulse phase that
are often coincident with pulses seen at X-ray and $\gamma$-ray
energies~\citep{lundgren1994}. \citet{popov2006b} propose that all
radio emission from the Crab (except for that in the precursor) is
composed entirely of GPs, consistent with the alignment of the GP and
high-energy components seen in other pulsars exhibiting GPs~\citep{cusumano2003,
  knight2006}.

The Crab pulsar shows pulsed emission across the entire
electromagnetic spectrum (see Fig.~\ref{fig:prof}, left), reflecting
different radiation processes in the pulsar magnetosphere~--- from
coherent curvature or synchrotron (radio) to incoherent synchrotron
(optical and X-ray) and incoherent curvature ($\gamma$-ray) radiation. Similar
to other sporadic variability phenomena seen in pulsar radio emission,
represented by nulling pulsars~\citep[e.g.][]{herfindal2009},
intermittent pulsars~\citep{kramer2006}, and rotating radio
transients~\citep{mmclaugh2006}, GP emission could be due to changes
in the coherence of the radio emission, variations in the pair
creation rate in the magnetosphere, or changes in the beaming
direction. If the GP phenomenon is due to changes in the coherence of the
radio emission mechanism, then one would expect little correlation of
the radio GPs with the high-energy emission.  However, if the GPs are
due to changes in the actual rate of pair creation in the pulsar
magnetosphere, one would expect an increased flux at high energies at
the time of the GPs. Similarly, because the radio GP and $\gamma$-ray
components are aligned, one expects that they come from the same place
in the pulsar magnetosphere. Therefore, if a GP occurs from a beam
direction alteration, one would expect to also see an increase in the
high-energy flux.

\citet{lundgren1995} previously attempted to carry out simultaneous
radio/$\gamma$-ray observations ($50-220$\,keV, the energy range of
CGRO/OSSE) and correlate times of arrival of GPs at 800 and 1300\,MHz with $\gamma$-ray photons. 
Their upper limit on the $\gamma$-ray flux
increase concurrent with radio GPs was $\le2.5$. Later,
\citet{ramanamurthy1998} correlated the same set of GPs with EGRET
photons of energy $>50$\,MeV, placing an upper limit on concurrent
$\gamma$-ray flux of 4.6 times the average Crab flux. This suggested
that the GP mechanism is largely based on changes in coherence and not
changes in pair production rates or beaming. Yet, \citet{shearer2003}
performed simultaneous radio/optical observations of the Crab pulsar
and found a weak correlation, i.e. that optical pulses coincident with
radio GPs were on average 3\% brighter than others. This observation
suggested that the GP emission mechanism, whatever its nature,
includes small variations in magnetospheric particle density.

\citet{lyutikov2007} proposed a more specific, quantitative model of
GP emission in which Crab GPs are generated on closed magnetic field
lines near the light cylinder via anomalous cyclotron resonance on the
ordinary mode. During emission of a photon, an electron undergoes transition {\it up}
in Landau levels.  The energy is supplied by the parallel motion
\citep{1985nypp.book.....G}. The application of anomalous cyclotron
resonance to pulsar radio emission has been discussed by
\citet{1999MNRAS.305..338L} and \citet{1979SvAL....5..238M}.

One clear prediction of this model is that radio GPs (at least those
at radio frequencies $>4$\,GHz) should be accompanied by $\gamma$-ray
photons, as the high energy beam is expected to produce curvature
radiation at energies $ \sim \hbar \gamma^3 \Omega \sim 0.1$--100\,GeV,
depending on the exact value of the Lorentz factor $\gamma$. These
energies fall into the energy range of the \textit{Fermi} mission, and so this
hypothesis can also be tested through high-frequency radio
observations concurrent with \textit{Fermi}.

The Fermi Large Area Telescope (LAT), with its large effective area,
broad field of view, and superior angular resolution is a perfect tool
for testing the Lyutikov theory and investigating the possible
correlation between GPs and $\gamma$-ray photons in general. For the
radio observations, using the 100-m Green Bank Telescope (GBT) allows one to
record a very large number of GPs within a reasonable observing time,
even at frequencies above 4\,GHz. Thus, a thorough study of the
correlation between high-energy $\gamma$-ray photons and
high-frequency GPs is possible.

In this paper, we present the results of simultaneous GBT/LAT
observations of the Crab pulsar. To probe the level of correlation
between GPs and $\gamma$-ray photons, we used two main approaches.
First, we searched for a ``burst correlation'' by examining whether
GPs cluster near $\gamma$-ray photons in time. Second, we analyzed
whether the average $\gamma$-ray flux of the pulsar increases within
the pulse phase windows where single GPs are produced.

In Sections~\ref{radioobs} and \ref{fermidata} below we describe the
radio observations and \textit{Fermi} data used in this analysis.
Section~\ref{scint} discusses the influence of the interstellar medium 
on the observed GP sample. We describe the correlation analysis
between radio GPs and \textit{Fermi} photons in Section~\ref{correlation}, and
conclude in Section~\ref{conclusions}.

\section{Radio observations}\label{radioobs}

\begin{figure}
\includegraphics[scale=0.43]{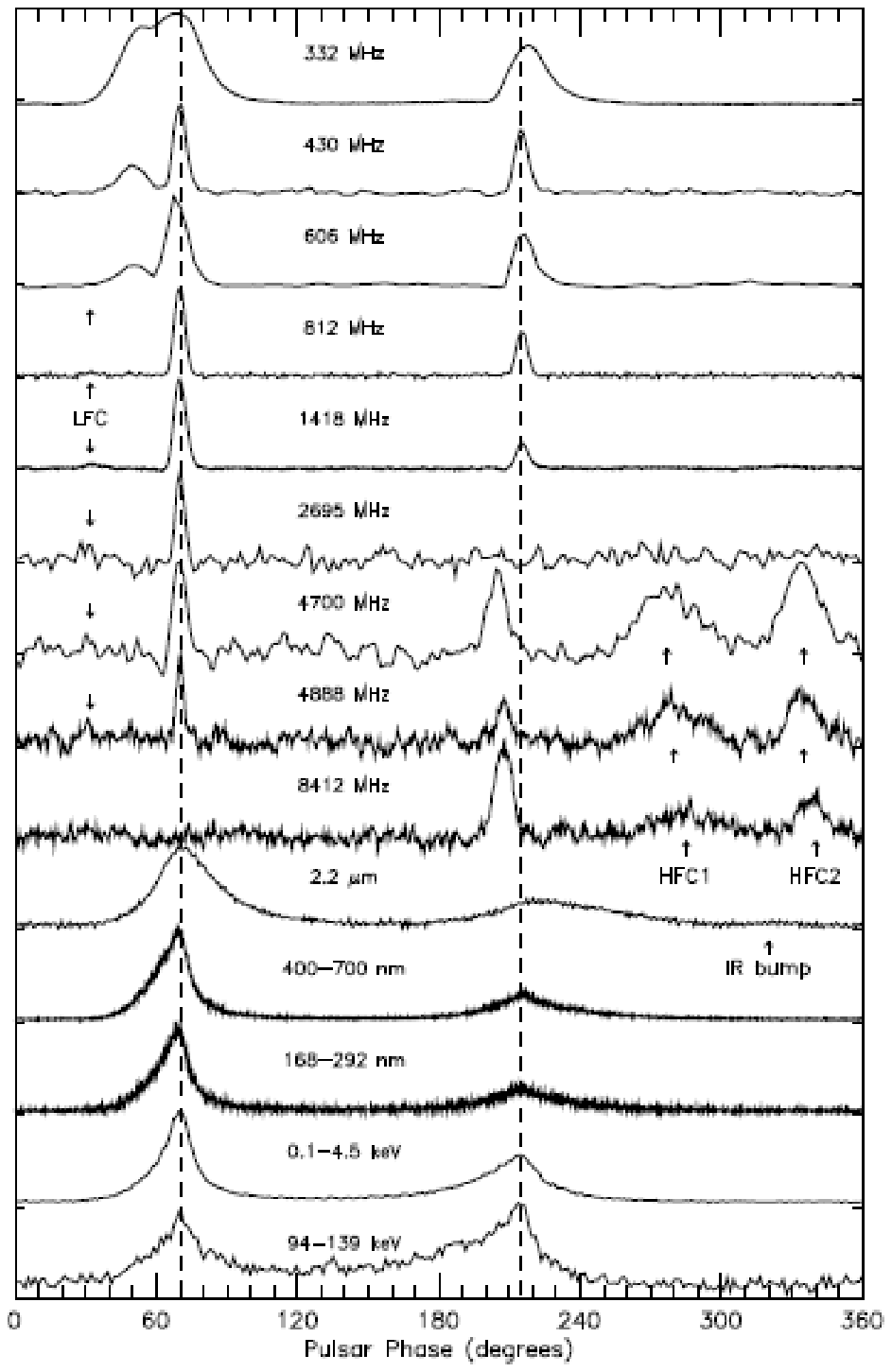}\includegraphics[scale=0.5]{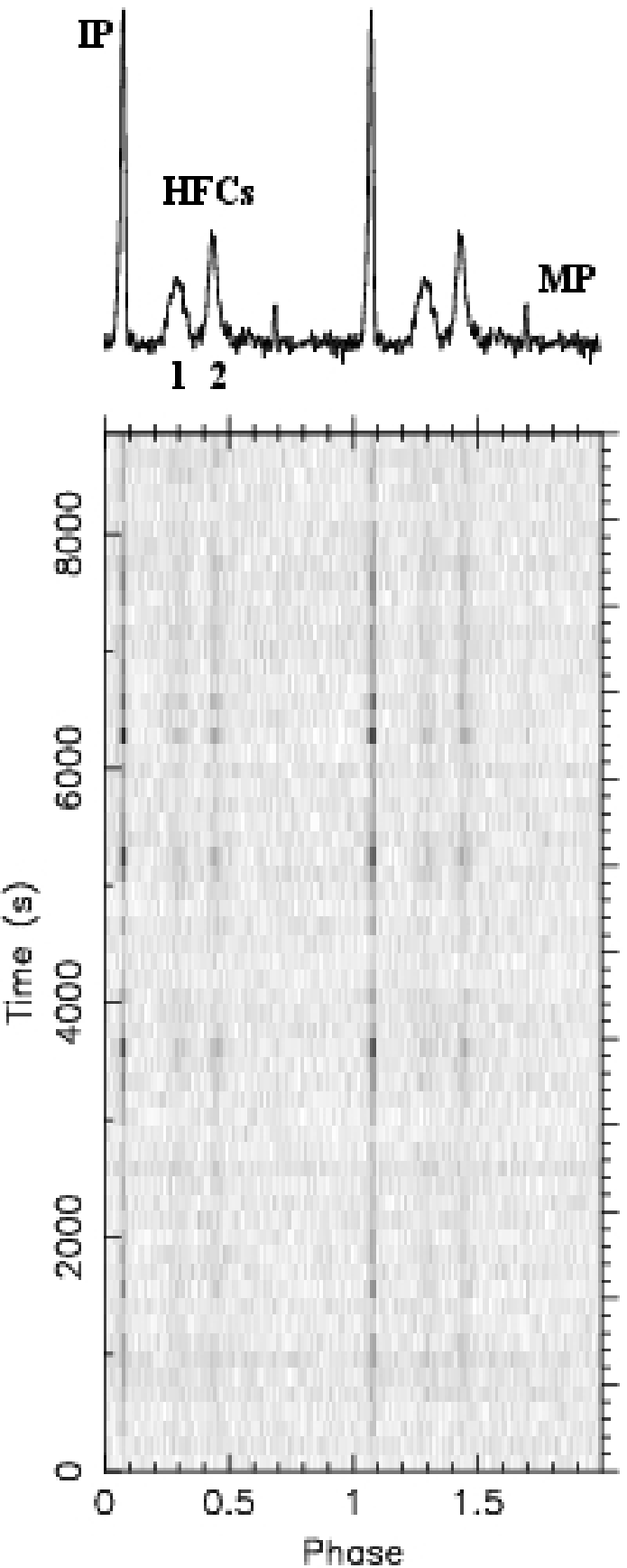}%
\caption{{\it Left.} Average profile of the Crab pulsar from radio to
  $\gamma$-rays, from the paper of \citet{moffett1996}. {\it Right.}
  Average Crab pulsar radio profile for one out of two sub-sessions with the GBT on Sep 25,
  2009.\label{fig:prof}}
\end{figure}

The radio observations were carried out during 12 observing sessions
in September--October, 2009 with the GBT, using the new Green Bank Ultimate Pulsar Processor Instrument
(GUPPI) at a central frequency of 8.9\,GHz, in incoherent dedispersion mode. The total bandwidth of 800\,MHz
was split into 256 frequency channels, and the total intensity was
recorded with a sampling interval 2.56$-$3.84\,$\mic$. Total observing
time was $\sim26$\,hrs or $\sim3\times10^6$ pulsar periods.

The raw data from every session were dedispersed with the current DM
of the Crab pulsar\footnote{The DM was 56.8005\,pc\,cm$^{-3}$ for
  September and 56.8109\,pc\,cm$^{-3}$ for October, from the Jodrell
  Bank Crab pulsar monthly ephemeris:
  \url{http://www.jb.man.ac.uk/pulsar/crab.html}} using
PRESTO package\footnote{\url{http://www.cv.nrao.edu/$\sim$sransom/presto/}},
and searched for all single-pulse events with signal-to-noise ratio (S/N) $>7$.  
Since GPs from the Crab pulsar do not have any established lower limit on peak flux density \citep{popov2006b}, we picked up 
initial threshold of S/N $>7$ in order not to contaminate our sample by 
numerous spurious detections on noise.
Each event was assigned a width, found by averaging the
dedispersed time series with different numbers of samples and finding
the number that resulted in a peak in S/N. The list of event times
was put into TEMPO2 format \citep{Tempo2} and converted to
the barycentric reference frame for the correlation analysis with
\textit{Fermi} data. Times of arrival (TOAs) were corrected for delay due to
propagation in the ionized interstellar medium (ISM).

Estimated timing errors due to an inaccurate DM are less than our time
resolution, assuming that DM varies smoothly and that between observing sessions the change in DM
 is less than the change over two months.  For 
$\mathrm{DM}_{\mathrm{Oct}} - \mathrm{DM}_{\mathrm{Sep}} =
0.0104$\,pc\,cm$^{-3}$, timing errors are about 0.5\,$\mic$.

Fig.~\ref{fig:prof} (right) shows the average pulse profile (top) of
the Crab pulsar at 8.9\,GHz together with the subintegrations from one of the
two sub-sessions on Sep 25, 2009, which had the highest rate of GP
detection of all 12 sessions. The interpulse (IP) and high-frequency components (HFCs)
are clearly seen, with the weak peak after HFC2 being the main pulse
(MP). However, on Sep~25 the pulsar was the brightest, and during
other sessions the average profile was less prominent.  During some
sessions we did not accumulate a detectable average profile at all.

\begin{deluxetable*}{ccccccccc}
\label{table_sum}
\tablecolumns{9}
\tablewidth{12cm}
\tablecaption{Summary of observational parameters and GP/$\gamma$-ray outcome
for each observing date. Columns include (from left to right): date of observation, time resolution,
system equivalent flux density (SEFD), total duration of radio observations and the time simultaneous with \textit{Fermi},
number of giant pulses, $N_{\mathrm{GPs}}$, detected during the whole observing session and during the time
simultaneous with \textit{Fermi}, number of $\gamma$-ray photons, $N_{\gamma}$.}
\tablehead{
\colhead{Date} & \colhead{$\Delta t$} & \colhead{SEFD} & \multicolumn{2}{c}{Radio time} & \colhead{} & \multicolumn{2}{c}{$N_{\mathrm{GPs}}$} & \colhead{$N_{\gamma}$} \\
\cline{4-5} \cline{7-8} \\
\colhead{}       & \colhead{}         & \colhead{}     & \colhead{Total} & \colhead{With \textit{Fermi}} & \colhead{} & \colhead{All~~} & \colhead{With}           & \colhead{} \\
\colhead{(2009)} & \colhead{($\mu$s)} & \colhead{(Jy)} & \colhead{(min)} & \colhead{(min)}               & \colhead{} & \colhead{}    & \colhead{\textit{Fermi}} & \colhead{}
}
\startdata
Sep 12 & 2.56\phn & 0.81\phn & 87.1\phn & 28.9\phn &   & 139\phn & 8\phn & 5\phn  \\
Sep 14 & 3.20\phn & 0.73\phn & 165.7\phn & 62.1\phn &  & 4375\phn & 1834\phn & 10\phn  \\
Sep 16 & 3.84\phn & 0.66\phn & 99.3\phn & 30.6\phn &   & 98\phn & 27\phn & 4\phn  \\
Sep 19 & 3.20\phn & 0.73\phn & 118.5\phn & 54.3\phn &  & 6957\phn & 1830\phn & 5\phn  \\
Sep 20 & 3.20\phn & 0.73\phn & 110.5\phn & 32.2\phn &  & 1846\phn & 384\phn & 2\phn  \\
Sep 21 & 3.84\phn & 0.66\phn & 55.1\phn & 31.5\phn &   & 27\phn & 16\phn & 2\phn  \\
Sep 22 & 3.20\phn & 0.73\phn & 147.7\phn & 68.9\phn &  & 1256\phn & 603\phn & 5\phn  \\
Sep 23 & 3.20\phn & 0.73\phn & 164.5\phn & 82.0\phn &  & 10520\phn & 5078\phn & 10\phn  \\
Sep 24 & 3.84\phn & 0.66\phn & 55.3\phn & 22.5\phn &   & 38\phn & 37\phn & 1\phn  \\
Sep 25 & 3.20\phn & 0.73\phn & 236.5\phn & 130.8\phn & & 14320\phn & 10014\phn & 13\phn  \\
Sep 28 & 3.20\phn & 0.73\phn & 72.3\phn & 48.1\phn &   & 34\phn & 9\phn & 5\phn  \\
Oct 25 & 3.20\phn & 0.73\phn & 157.6\phn & 41.3\phn &  & 3164\phn & 1261\phn & 15\phn  \\
\noalign{\smallskip}\hline\noalign{\smallskip}
Total & & & 1470.0\phn & 633.1\phn & & 42774\phn & 21092\phn & 77\phn
\enddata
\label{table_sum}
\end{deluxetable*}

The system equivalent flux density (SEFD) is mostly determined by the
Crab Nebula. Flux densities for the Crab Nebula were calculated with the relation
$S(f) = 955\times(f/\mathrm{GHz})^{-0.27}$\,Jy \citep{cordes2004},
accounting for the fact that at 8.9\,GHz the solid angle of the GBT
beam covers only 25\% of the area occupied by the nebula.  We estimate
a SEFD~$= 1.3 /\sqrt{\Delta t / 1 \mic}$\,Jy, or about 0.7\,Jy for our
most common sampling time, or $\Delta t$, of 3.2\,$\mic$.

Preliminary analysis of all events with S/N $>$ 7 revealed that GPs
appear mostly in the MP and IP phase windows. Only for 2 observing
sessions, namely on Sep 25 and 28, were there several GPs detected in the
HFCs, all with S/N close to 9. This is to some extent surprising,
since \cite{jessner2005} observed the Crab pulsar with similar parameters and
timespan, and found about 120 GPs in HFCs versus 350 GPs in the MP and IP. However,
their threshold peak flux density of 25\,Jy was much higher than our threshold of about 6\,Jy, 
suggesting that the GPs from HFCs are rarer but brighter.

\begin{figure*}[h]
\includegraphics[scale=0.68]{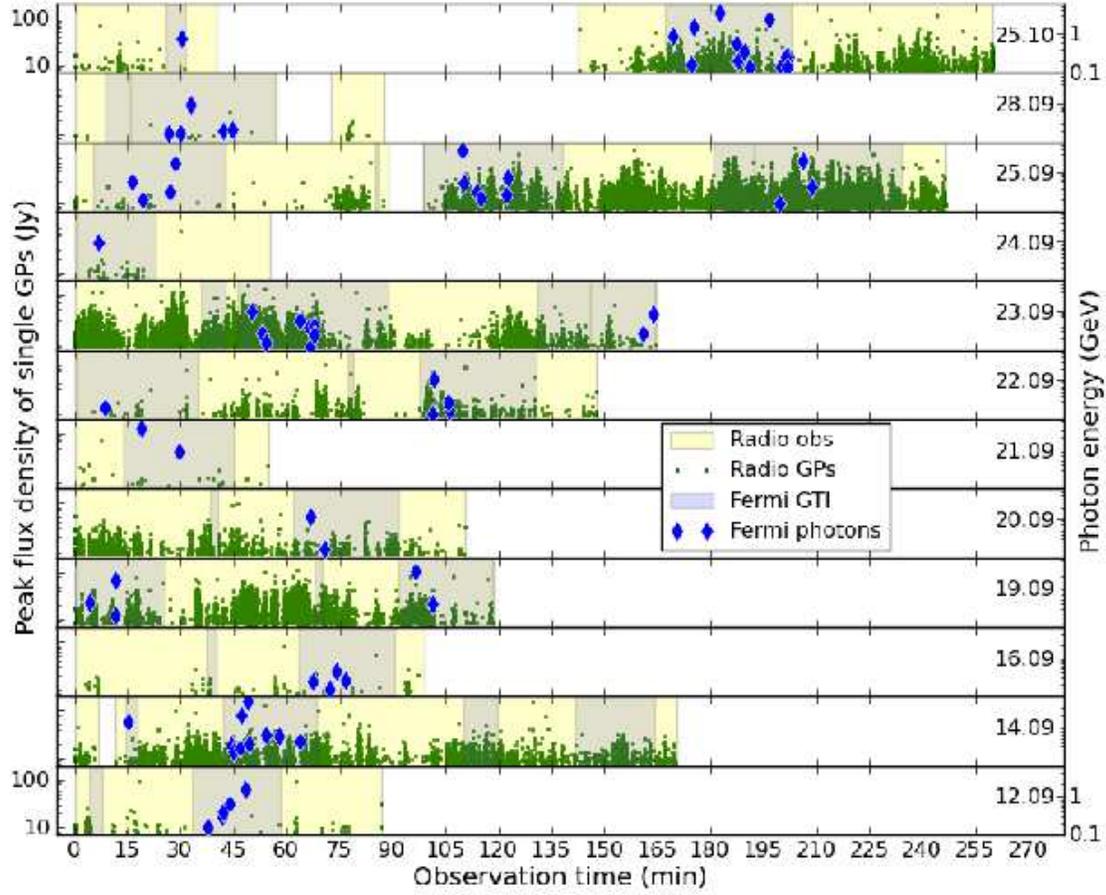}
\caption{Time series of radio GPs and \textit{Fermi} photons during 12
  observing sessions. X-axis -- time from the beginning of each
  session, in minutes. Y-axis (left) -- peak flux density of radio GPs.
  Y-axis (right) -- energy of $\gamma$-ray photons. Both scales are the same for each observing session. The yellow shaded regions mark
  the time when we actually were recording radio data and the blue are
  the \textit{Fermi} Good Time Intervals. The observing date is given in 
  the upper right-hand corner of each subplot. For the observing session on Oct~25 two photons came within short time interval, so their markers overlap and one can see only 14 photons, instead of 15. }
\label{fig:burst}
\end{figure*}
\begin{figure*}[h]
\centering
\includegraphics[scale=0.4,angle=0]{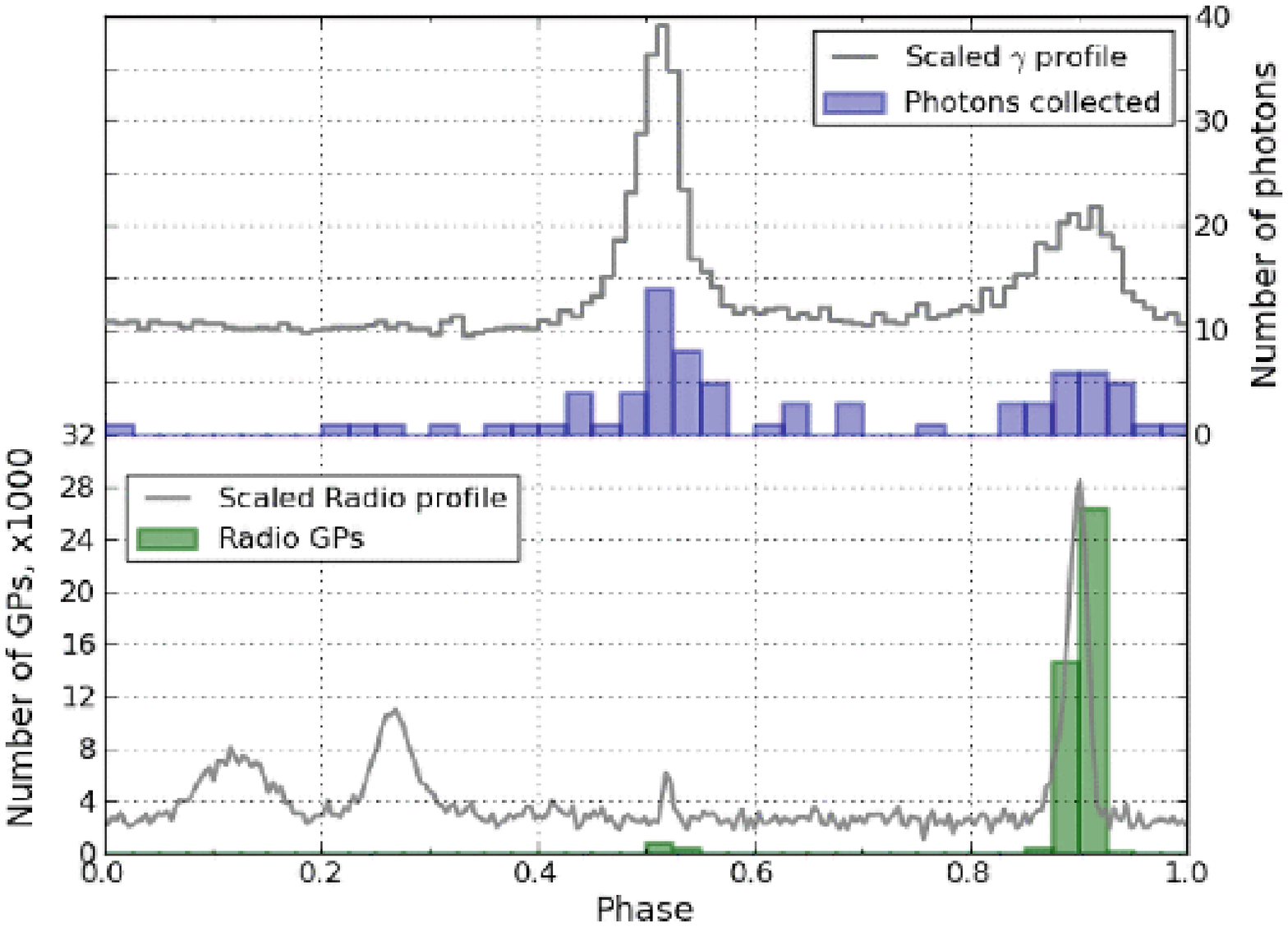}\includegraphics[scale=0.4,angle=0]{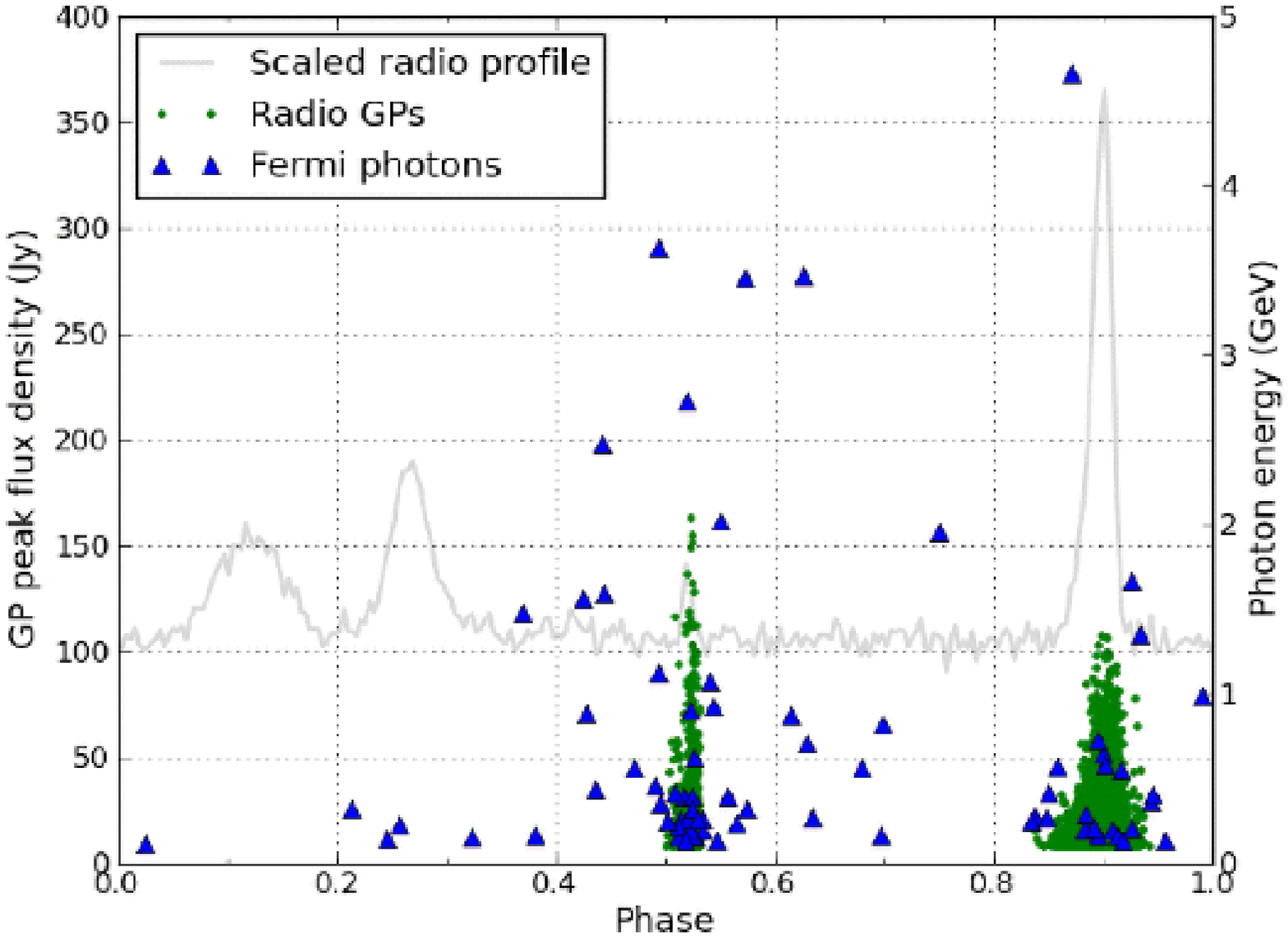}
\caption{\textit{Left:} histograms of GPs for all radio observing time (bottom) and \textit{Fermi} photons during the simultaneous time (top). For illustrative purposes, scaled radio and $\gamma$-ray profiles (grey) are shown with arbitrary offset along y-axis.  The scaled radio profile is from the second scan of GBT session on Sep~25, 2009 (as shown in Fig.~\ref{fig:prof}), and the $\gamma$-ray profile is the \textit{Fermi} profile accumulated during Sep-Oct, 2009. \textit{Right:} distribution of peak flux density of GPs and energy of $\gamma$-ray photons over pulsar rotational phase. Also, there is shown the scaled radio profile as for histogram on the left.}
\label{fig:profhist}
\end{figure*}

Some of our sessions were heavily contaminated with broad radio
frequency interference (RFI) pulses with typical S/N $<$ 10. Therefore,
we analyzed only events with a peak flux density exceeding
8.1\,Jy (S/N = 10 for the session with smallest sampling
time) and which arrived in the MP or IP phase windows.  Additionally,
we excluded all events with width larger than 30 sample intervals, as
being presumably caused by RFI. These cuts resulted in the selection
of more than 40000 GPs. 
 Comparing the
number of single pulses above 8.1\,Jy and narrower than 30 samples in
and out of the pulsed emission phase windows, we can estimate the
proportion of \textit{false GPs} in our final data set to be less than
0.001\%.

The summary of observations is given in Table~\ref{table_sum}. For each observing  date the listed
columns are: time resolution, SEFD, total duration of radio observations and the time simultaneous with \textit{Fermi}, 
number of giant pulses, detected during the whole observing session and during the time
simultaneous with \textit{Fermi}, and the number of $\gamma$-ray photons selected for further analysis (see Section~\ref{fermidata}).

\section{\textit{Fermi} data}\label{fermidata}

We extracted ``Diffuse" class events  with energies from 100\,MeV to 300\,GeV 
from the \textit{Fermi} database, concurrent with each radio observation. Photons with zenith
angles greater than $105^{\circ}$ were excluded to eliminate the $\gamma$-rays generated in the Earth's atmosphere.

We selected only photons in Good Time Intervals (GTIs) within an angle
$\theta < \mathrm{Max}(6.68 - 1.76\log(E/1000~\mathrm{MeV}),1.3)
^\circ$ of the radio pulsar position \citep{FermiCrab2010}. Photon arrival times 
were converted to the Solar System barycenter and assigned phases with the TEMPO2
\texttt{fermi} plugin. The timing accuracy of the \textit{Fermi} LAT is better
than 1\,$\mic$ \citep{FermiCrab2010}. LAT dead time per event is less
than
100\,$\mic$\footnote{The dead time  was taken from \textit{Fermi} Technical Handbook, \\ \texttt{http://fermi.gsfc.nasa.gov/ssc/proposals/manual/}}, which is
less than 3\% of pulsar rotational phase.  Over the course of all
radio observations we accumulated 10.5\,hours of \textit{Fermi} data within
GTIs, resulting in 77 photons with energies above 100\,MeV (see Table~\ref{table_sum}).

Fig.~\ref{fig:burst} gives a quick visual summary of our simultaneous
observations, showing \textit{Fermi} photons and radio GPs versus observing
time for each session. The distribution of number and energy/peak
intensity of photons/GPs with respect to pulsar rotational phase is
shown in Fig~\ref{fig:profhist}. As reported earlier, $\gamma$-ray and
radio emission windows are aligned.

\section{Propagation effects for Giant Pulses} \label{scint}

\begin{figure*}[tb]
  \centering
  \includegraphics[scale=0.44]{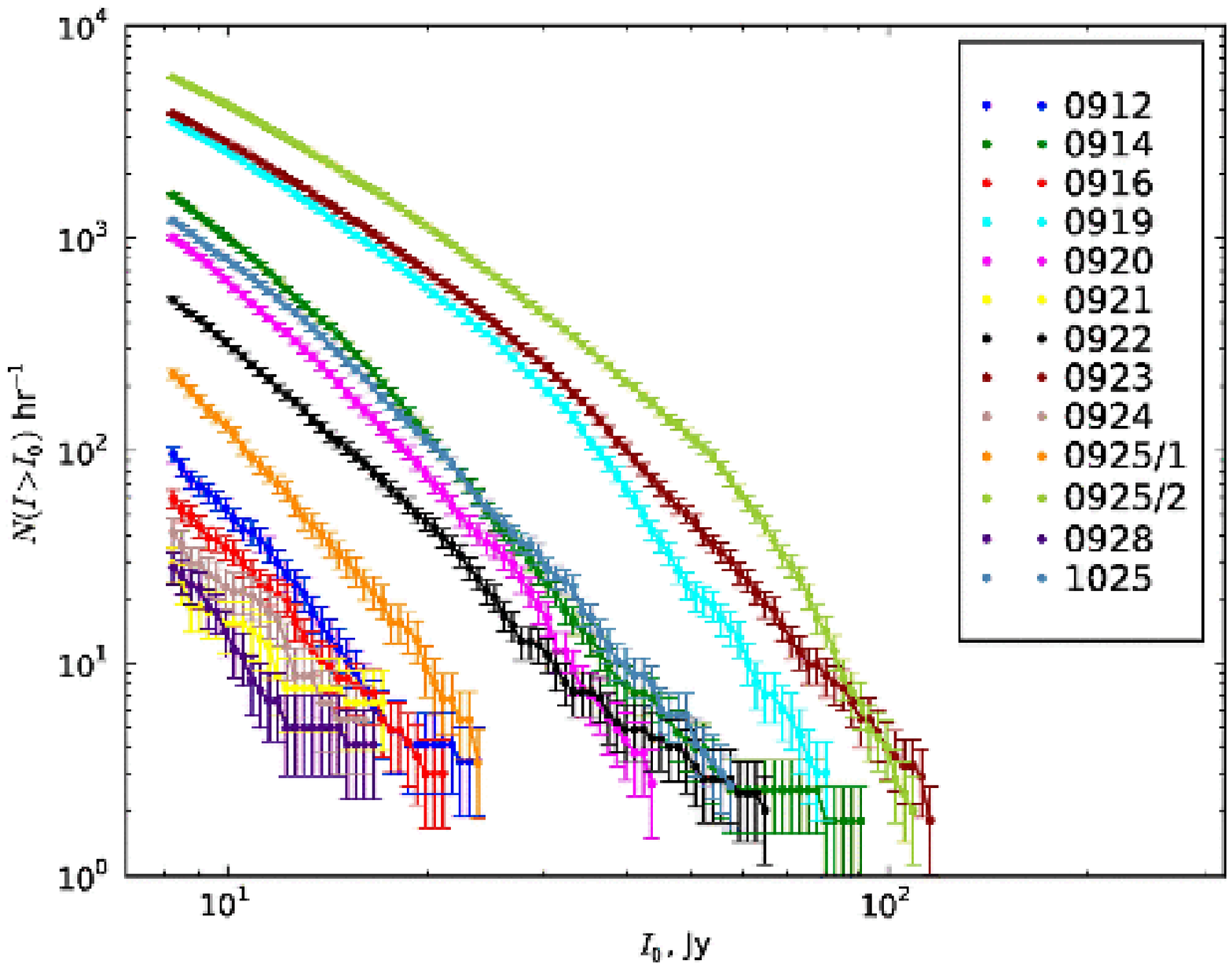}
  \includegraphics[scale=0.44]{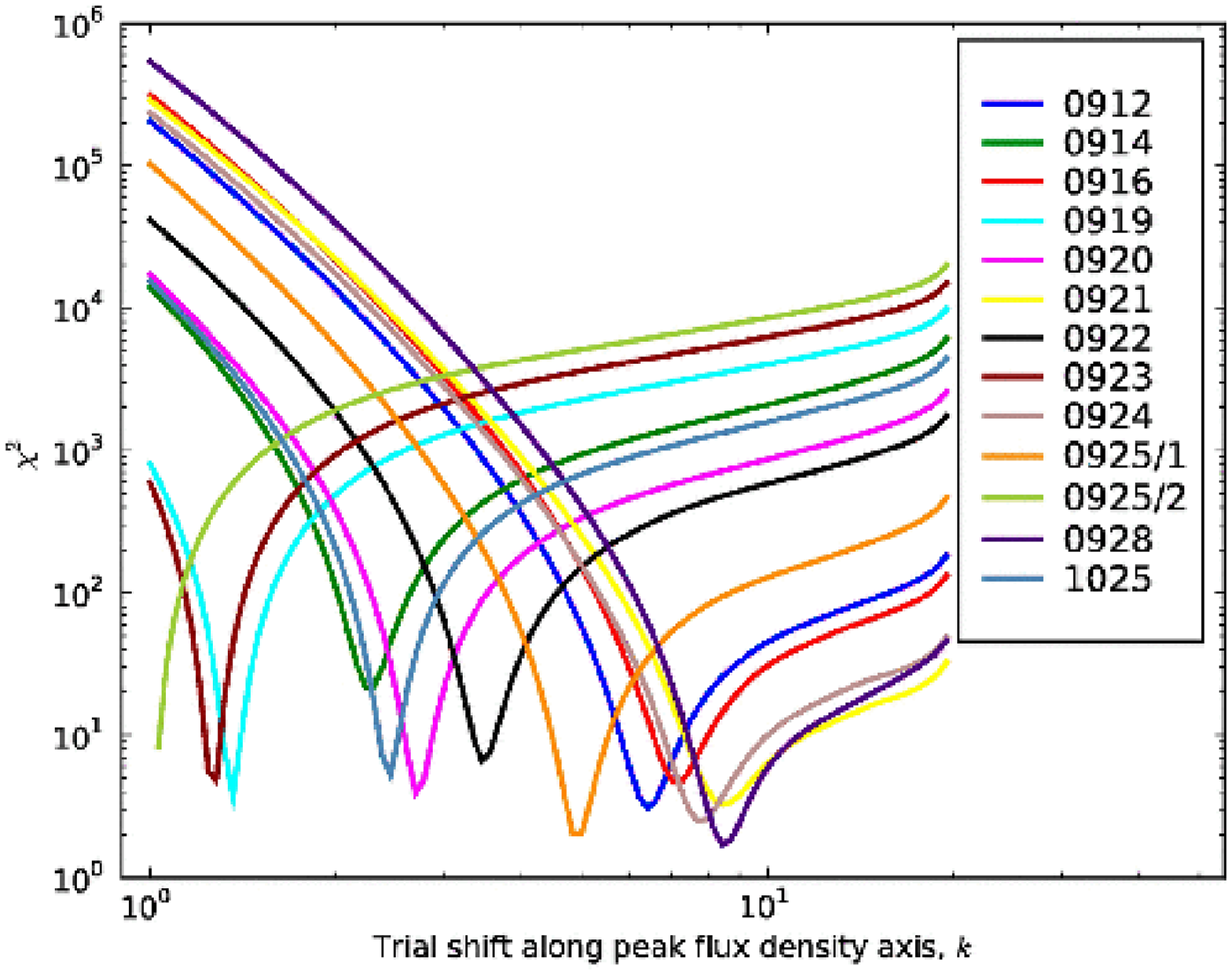}
  \caption{\textit{Left}: cumulative peak flux distribution of GPs (MP
    and IP together) for different observing sessions and 2
    subsessions of Sep~25 with apparently different GP rates. Poisson errors are also shown. \textit{Right}: $\chi^2$ values versus the RISS damping coefficient, $k$, for each observing session.}
  \label{fig:k}
\end{figure*}
At high frequencies, GPs are strongly affected by interstellar scintillations \citep{cordes2004}, 
which change their apparent rate and peak intensities. Careful treatment of ISM effects is
crucial for a proper correlation analysis.  Unfortunately, we did not
make direct measurements of typical ISM diagnostic parameters, such as
scintillation bandwidth and pulse broadening time.  Instead, we use
the scintillation timescales scaled from other frequencies.  These values give only a rough estimate of scintillation timescales at 8.96\,GHz, since for the Crab pulsar the
main contribution to scintillations is made by the turbulent and
quickly changing Crab Nebula. Scintillation parameters for the Crab
have been shown to be strongly variable with time (see
\cite{cordes2004} and references therein).

The refractive interstellar scintillation (RISS) timescale, $\tau_{\mathrm{RISS}}$, can be scaled using the $\nu^{-2.2}$ dependence derived from a five frequency
data set \citep{rickett1990}. At 8.96~GHz, $\tau_{\mathrm{RISS}}$ is
about 80~minutes, thus roughly matching the observed day-to-day
variation of the GP rate (see Fig.~\ref{fig:burst}). Also, for the longest, 4-h session,
note the change of GP rate between two 2-h scans in Fig. \ref{fig:burst} (observing settings were the same for both scans).

Following \cite{cordes2004}, for calculating the diffractive
interstellar scintillation timescale, $\tau_{\mathrm{DISS}}$, we
adopted the thin screen model with a Kolmogorov spectrum of
irregularities and reference pulse broadening time $\tau_d = 0.5$\,ms
at 0.3\,GHz. At 8.9\,GHz, with bandwidth of 800\,MHz, this gives us a
scintillation strength $u = 8.9$, well into the strong scintillaton
regime, and $\tau_{\mathrm{DISS}} = \tau_{\mathrm{RISS}}/u^2 \simeq
9$\,min.

The variation of the GP rate within each observing subsession was
estimated by autocorrelating the rate of GP emission in 10-second
bins. The autocorrelation analysis shows two shorter GP rate
variability time scales of $\sim$20~minutes and 2$-$4~minutes. Both of
them agree fairly well with the DISS estimations, considering all the
uncertainty in the scintillation parameters. On the other hand, there
is no evidence against the hypothesis that at least one of these
timescales is due to intrinsic GP rate variability.

\textit{The following analysis assumes that observed day-to-day variation 
(or, in case of Sep 25, variation between 2 subsessions) of the GP rate and mean intensity is caused by RISS.} If \textit{intrinsic} GP rate and mean intensity are
constant on timescales larger than 90\,min, then it is relatively easy to make a GP sample corrected for refractive scintillation. We accomplish this by
estimating the amount of RISS intensity damping on each observation session 
with respect to the session with the highest GP rate.  Then, we multiplied the intensities of all pulses in each separate session
by those amounts, and threw out all GPs below a threshold, common for
the corrected GPs over all sessions. 

A simple way to calculate the intensity variation due to RISS would be
by comparing mean profiles of pulsed emission accumulated during each
session.  However, at these frequencies, our observations were not
sensitive enough to accumulate the normal Crab pulse profile except
on one or two sessions where scintillations caused a boosting of the
average flux density of the pulsar.  Instead, we compared the
intensity distributions of GPs between sessions.  If the change in
rate and mean intensity of GPs on timescales of a few hours is due to
RISS, then the peak intensity distributions for each day should have
the same shape, but with different values of peak flux density.  The
distributions in Fig.~\ref{fig:k} (left) show that this assumption is
basically correct.

\begin{table}[t]
\begin{center}
\caption{RISS correction coefficients $k$ for each observing session.}
\begin{tabular}{lcccccc}
\hline
Session  & RISS correction coefficient $k$ & $\chi^2_{\mathrm{min}}$  \\
\hline
Sep 12 & 6.4 & 3.1 \\
Sep 14 & 2.2 & 21.7 \\
Sep 16 & 7.2 & 4.7 \\
Sep 19 & 1.4 & 3.8 \\
Sep 20 & 2.7 & 3.9 \\
Sep 21 & 8.5 & 3.3 \\
Sep 22 & 3.4 & 6.6 \\
Sep 23 & 1.3 & 4.8 \\
Sep 24 & 7.7 & 2.5 \\
Sep 25/1 & 5.0 & 2.0 \\
Sep 25/2 & 1.0 & 0.0 \\
Sep 28 & 8.5 & 1.7 \\
Oct 25 & 2.5 & 5.2 \\
\hline
\end{tabular}
\label{table:k}
\end{center}
\end{table}

As the reference session, we picked the one with the highest rate of
GPs, the second subsession of Sep 25 (from now on called ``0925/2''
or ``reference session''). For each day and for the two subsessions on
Sep 25 separately, we determined the RISS damping coefficient, $k$, by
minimizing $\chi^2$:
\begin{equation}
\chi^2_k =
\frac{1}{N_{\mathrm{bins}}-1}\sum_{I_i}\frac{\left[N_{\mathrm{ref}}(I_i) -
    N(I_i/k)\right]^2}{\sigma^2_{N(I_i/k)}},
\end{equation}
\noindent
where $N(I_i)$ is the number of GPs with peak intensity higher than
$I_i$, per hour of observation. $N_{\mathrm{bins}}$ is the number of
bins in the distributions being compared, and $\sigma_{N(I_i/k)} =
\sqrt{N(I_i/k)/T_{\rm hrs}}$ is the Poisson error in each bin if
$T_{\rm hrs}$ is the duration of the session in hours and assuming
that the energy of each GP does not depend on the energy of the
preceding one.

\begin{figure*}[t]
  \centering
  \includegraphics[scale=0.45]{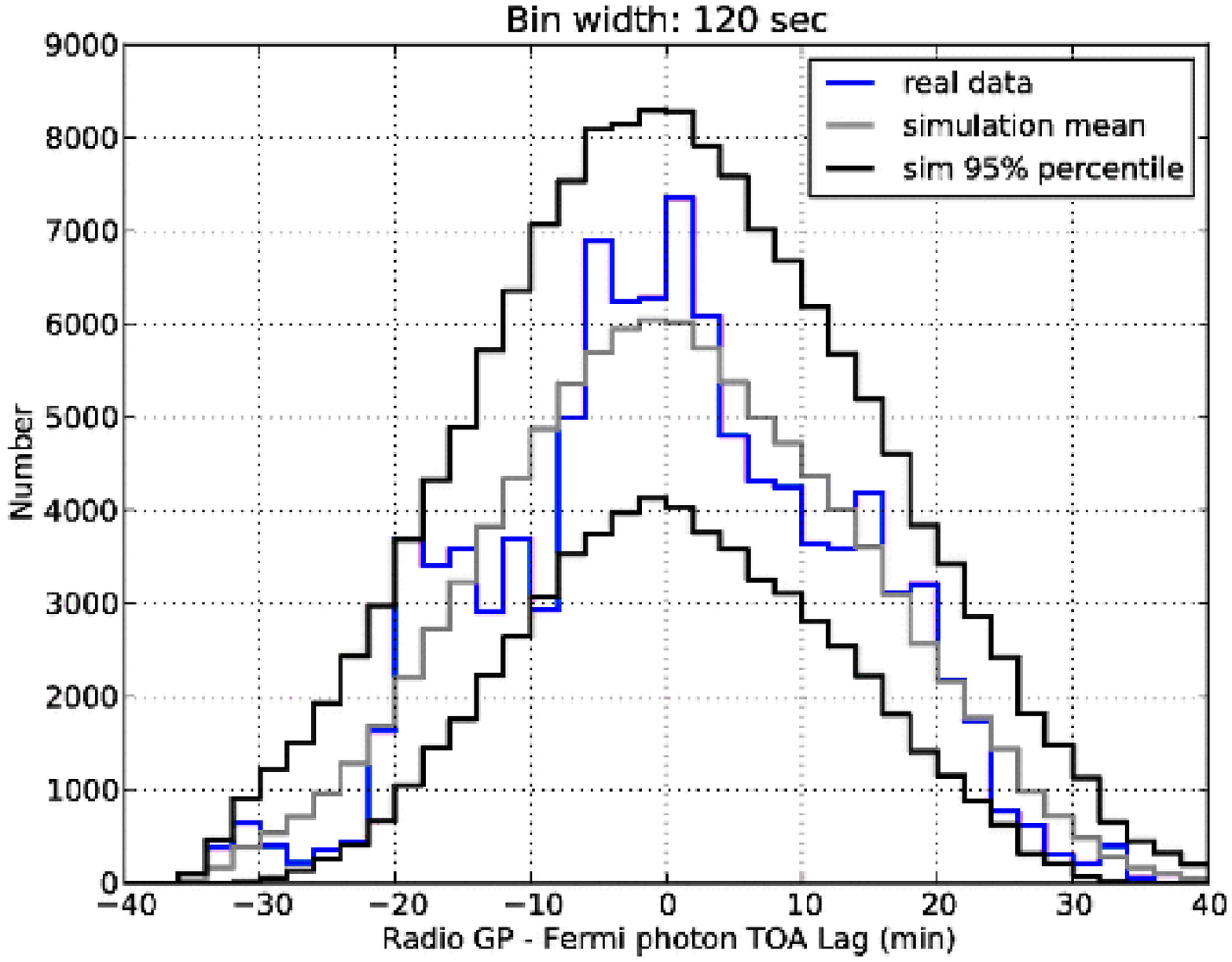}\includegraphics[scale=0.45]{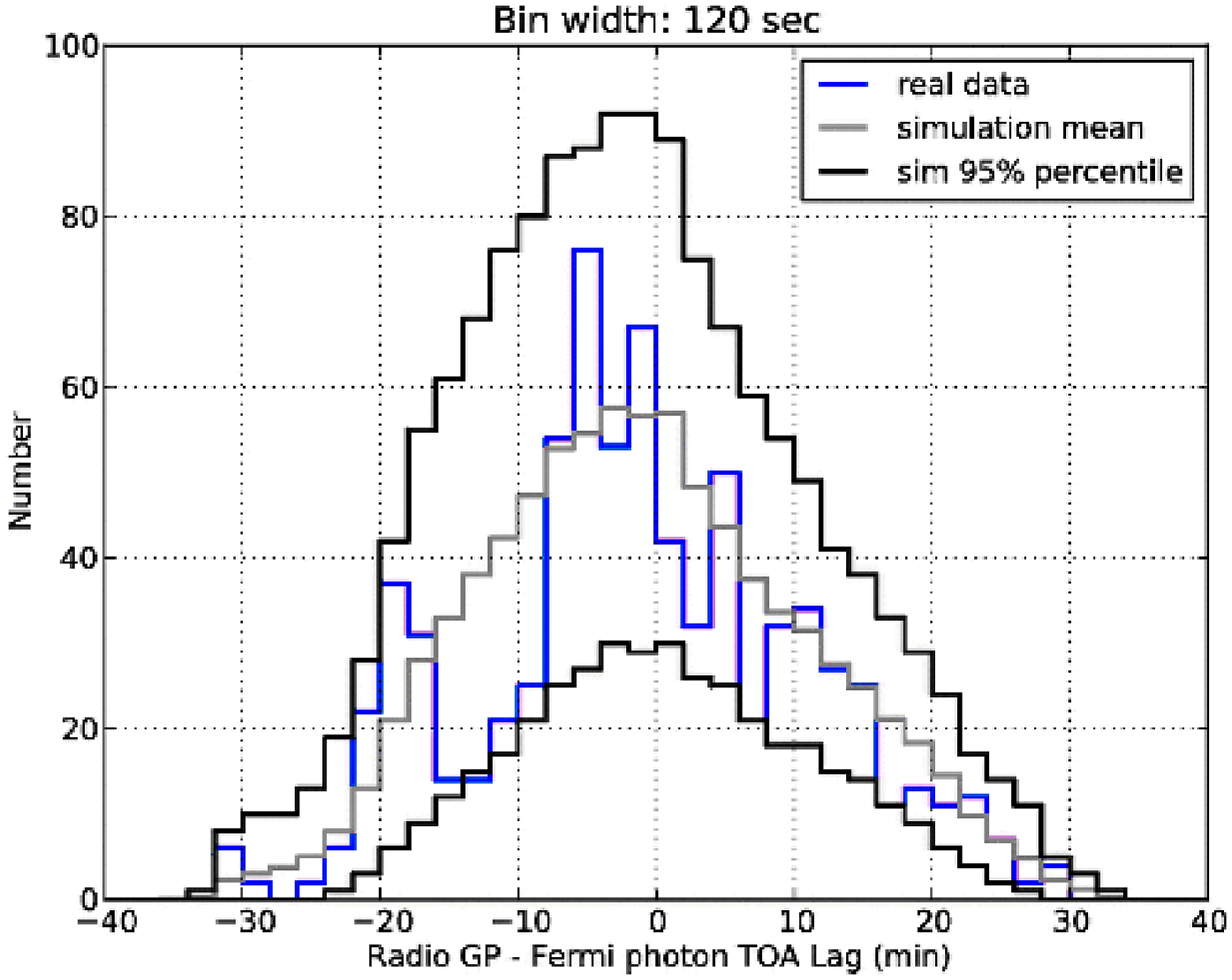}
  \caption{The distribution of time lags between GPs and $\gamma$-ray
    photons for 2-min bins, for all GPs $> 8.1$\,Jy (21000 GPs; {\em
      left}) and the RISS-corrected GPs 
    (180 GPs; {\em right}). Real \textit{Fermi} photons
    (blue line) are contrasted to the mean and 95\% percentile on the
    pool of simulated data sets (grey and black). The fact that the
    distribution for the real data lies within 95\% of the simulated
    ones indicates no apparent change in GP generation rate on 2-min
    timescale with any possible time lag up to $\pm 40$ minutes with respect to the $\gamma$-ray photons. The maximum lag value, 40\,min, corresponds to the size of the largest GTI window. All other bin widths (down to 10\,s) give the same result.}
  \label{fig:correlation-burst}
\end{figure*}

The $\chi^2_k$ curves are plotted in Fig.~\ref{fig:k} (right). All but
one have a sharp minimum of $\chi^2_k \lesssim 10$, indicating
reasonable fits, at $k$s between 1 and 10. The only outlier is the
session of Sep 14th, with $\chi^2_{\mathrm{min}} = 21.7$, which has an
abnormal excess of intrinsically strong GPs (see Fig.~\ref{fig:burst}).  These GPs
do not exhibit any other peculiar properties, other than relatively high peak
flux density.

Both RISS coefficients and the corresponding $\chi^2_{\mathrm{min}}$
for each day are listed in Table~\ref{table:k}. We corrected the
session GPs by multiplying their flux densities by $k$ and set the
intensity threshold for corrected pulses as 8.1\,Jy$\times \max(k) = 69$\,Jy (8.1\,Jy was our initial threshold, see Section \ref{radioobs}).

Thus, we effectively selected only those pulses which would have had
peak flux density larger than 8.1\,Jy \textit{if} they were observed during the session 
with highest RISS damping. Our uniform sample of such intrinsically brightest
pulses numbered 180 GPs with TOAs within the \textit{Fermi} observing time.

\section{Correlation analysis}\label{correlation}

The correlation analysis focused on two distinct tasks. The first one
aimed to probe if the GP generation rate correlates with observed
$\gamma$-ray photons. For the second, we investigated the hypothesis of
\cite{lyutikov2007} that the $\gamma$-ray photon flux increases during
GPs. For both cases, we used simulated high-energy data sets with no
assumed intrinsic correlation between the GPs and $\gamma$-ray photons to
test the statistical level of correlation present in the real data.

The simulations used the \texttt{gtobssim} software from the \textit{Fermi}
tools package. We used the latest version of instrument response
function, Pass6\_v3 together with the same spacecraft/pulsar ephemeris
as in real data analysis.

We simulated the pulsar using the \texttt{PulsarSpectrum} library,
with the light curve, spectrum and integral flux above 100\,MeV taken
from \cite{FermiCrab2010}. The integral flux was set to
$F_{\mathrm{av}} = 2.09\times 10^{-6}\,\mathrm{cm^{-2}s^{-1}}$ for the
burst correlation analysis and varied from 0 up to about a hundred
$F_{\mathrm{av}}$ for the single-pulse correlation analysis (see~\ref{correlation-sp}). When we
simulated zero flux from the Crab pulsar, we simply removed the pulsar
from the list of simulated sources.

We modeled the Crab Nebula as a point source (for the energy ranges in
question its angular diameter is less than the {\em \textit{Fermi}} region of
interest), with the spectrum as determined in \cite{FermiCrab2010} and
integral flux above 100~MeV of $9.8\times 10^{-7}\,\mathrm{
  cm^{-2}s^{-1}}$.  For the Galactic and extragalactic backgrounds we
used the ``GalacticDiffuse\_v02'' and ``IsotropicDiffuse\_v02''
models.  The simulated photon files were processed in the same way as
the real data.

\subsection{Is GP rate correlated with single $\gamma$-ray photons?}\label{correlation-bursts}

To test if $\gamma$-ray photons are correlated with the GP generation rate,
we calculated the distribution of time lags between each photon and
all GPs in that photon's GTI. The same procedure was applied to the
simulated $\gamma$-ray data sets, such that if there were any clustering of
GPs around $\gamma$-ray photons (or with some time lag with respect to the
$\gamma$-ray photons), it would be seen as a discrepancy between the
real and simulated distributions of the high-energy data. Changing the
bin size of the distribution makes it sensitive to different
timescales of possible clustering of GPs. In this study, we tried a
set of bin widths, starting from 10 seconds and increasing the width
by 10 seconds up to 2 minutes. Two minutes corresponds to the smallest
timescale of GP clumping (likely caused by interstellar
scintillation), as shown in Fig.~\ref{fig:burst}.  On timescales less
than 10\,s, the Poisson noise due to a discrete number of time lag measurements
becomes too high. We performed 1000 simulation runs and
contrasted the real-data distribution with the mean and 95\%
percentile of all of the simulated data sets.

Fig.~\ref{fig:correlation-burst} shows the distribution of time lags
between GPs and photons for all GPs and for a RISS-corrected
sample of GPs for one particular bin width, namely, 2\,min.  In both
cases the real data set lies all within the 95\% percentile of the
simulations, indicating no apparent change in GP generation rate on
2-min timescales with any possible time lag within $\pm 40$ minutes (maximum GTI length) of
the arrival of the $\gamma$-ray photons. All other bin widths, down to 10\,s, gave the
same result.

\subsection{Does $\gamma$-ray flux change around single GPs?}\label{correlation-sp}

Another question of interest is whether the average $\gamma$-ray flux
from the Crab pulsar increases during individual giant pulses, as
predicted by \cite{lyutikov2007}.  To investigate that, we looked for
the number of $\gamma$-ray photons in on-pulse emission windows around
each GP. We performed a separate search for all GPs, looking for
photons in a large window consisting of the main pulse, interpulse and
bridge between them, and also for IP GPs only, limiting the
correlation window to the interpulse phase range.

\begin{figure*}[t]
  \centering
  \includegraphics[scale=0.45]{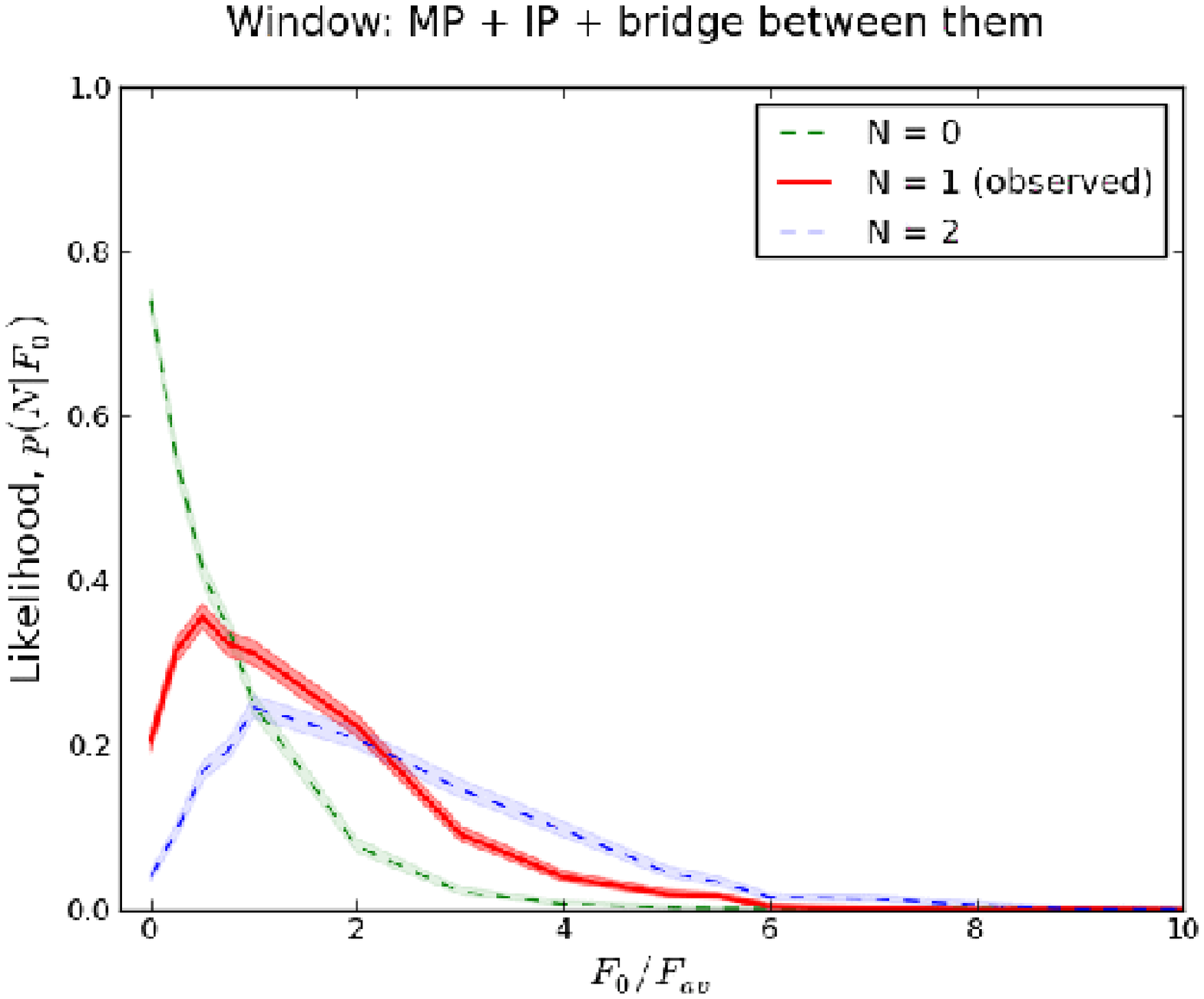}\includegraphics[scale=0.45]{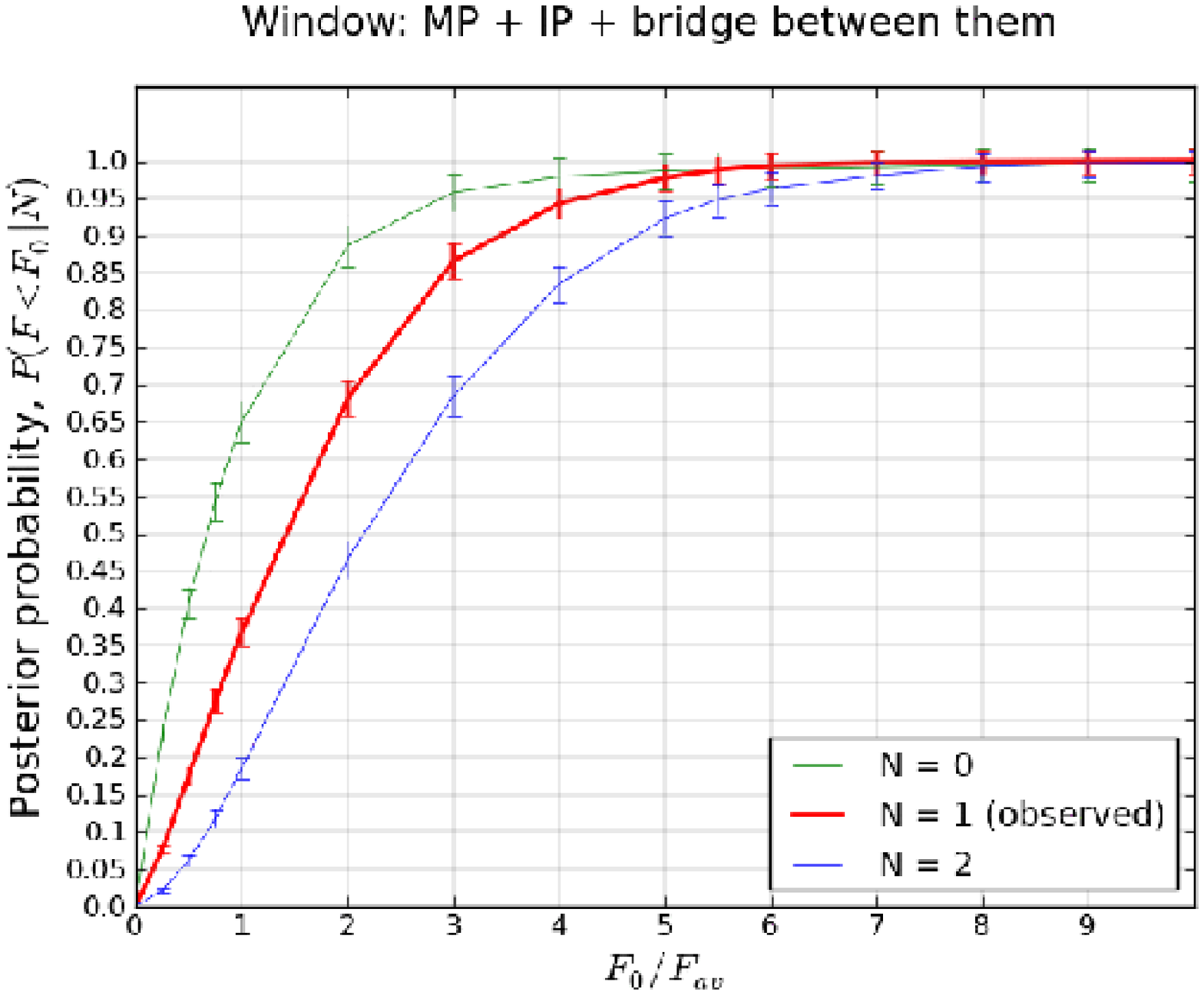}
  \includegraphics[scale=0.45]{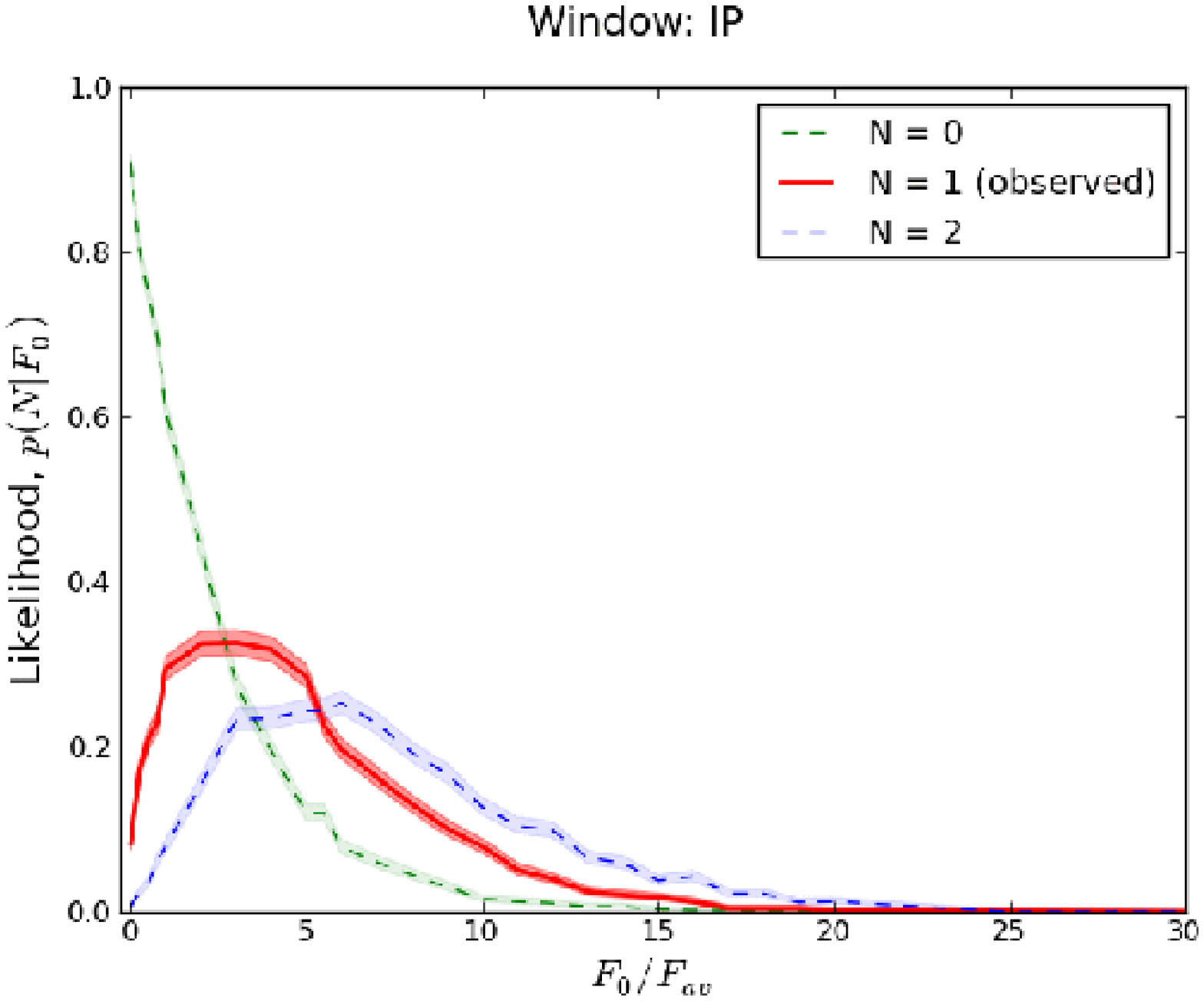}\includegraphics[scale=0.45]{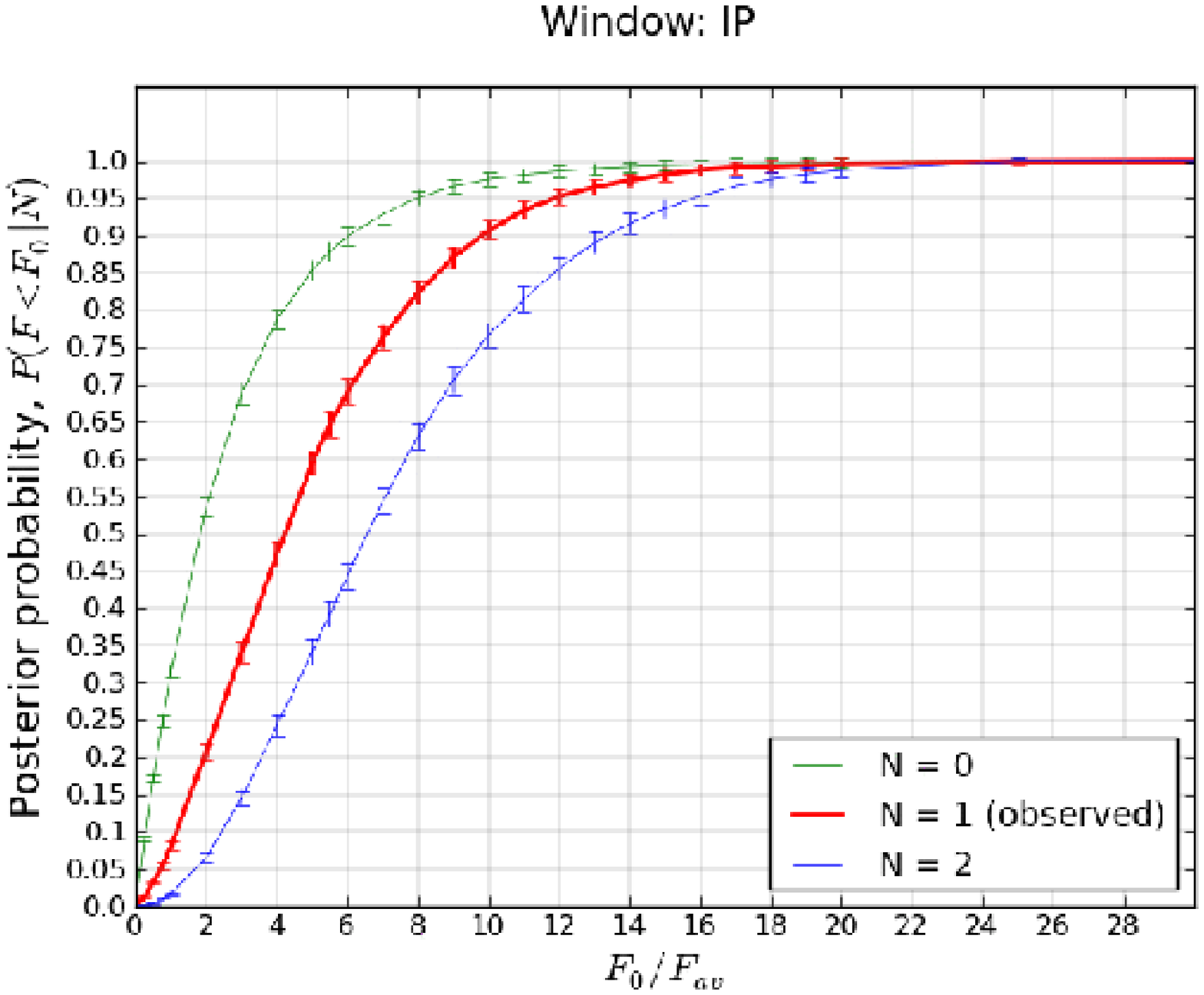}
  \caption{\textit{Left}: Likelihood, the probability of
    observing $N$ photons in a certain window around a GP if the
    $\gamma$-ray flux from the pulsar in this window were $F_0$. For
    each window size, we observed only one such match between GP and
    $\gamma$-ray photon, so $N=1$. Since photon arrival times are a
    Poisson process, the error on $N$ is $\sqrt{N} = 1$, thus
    likelihood curves for $N=0$ and $N=2$ are also shown. The shaded
    region around each curve corresponds to uncertainty due to the
    limited number of simulation runs (see text for explanation). Both $N=1$ 
    likelihood curves peak near $F_0/F_{\mathrm{av}}=1$, so no or weak 
   correlation between GPs and $\gamma$-ray photons is the most probable. 
    \textit{Right}: Posterior probability that $\gamma$-ray flux in a
    window around GP is \textit{less} than $F_0$, given the observed
    number of matches $N$. Errors due to the limited number of
    simulation runs are plotted as errorbars, whereas those due to a
    discrete number of matches are given by separate posterior
    probability curves for $N=0$ and $N=2$. $F_{\mathrm{av}}$, the
    average pulsed $\gamma$-ray flux from Crab pulsar, is from
    \cite{FermiCrab2010}.}
\label{fig:correlation-bayes}
\end{figure*}


If a photon was detected in a window around a GP, it was called a
``match''. For 10.5~hours of simultaneous observations, \textit{we
  detected only one such match}: a photon with $E=403.7$\,MeV was detected within 1.3\,ms of IP GP with peak flux density of 8.9\,Jy. Nonetheless, knowing the observed number of
matches $N=1$, the probability that the $\gamma$-ray flux during GPs
is equal to some value $F_0$ can be estimated with the simple Bayesian
formula:
\begin{equation}
p(F = F_{0} | N) = \frac{p(F_0)\cdot p(N | F =
  F_{0})}{\int_0^{F_{max}} p(F_0) \cdot p(N | F = F_{x})dF_x}
\label{eq:bayes}
\end{equation}
where $p(F_0)$ is the prior distribution for $F_0$ and $p(N | F =
F_{0})$ is the likelihood, i.e. the probability to get the observed
number of matches $N$ if the pulsar $\gamma$-ray flux during GPs is equal to $F_0$.

Since little is known about $p(F_0)$, the prior distribution for
$F_0$, we chose the prior to be uniform in a flux range from 0 (the
Crab pulsar turns off $\gamma$-ray emission during GPs) to
\begin{equation}
F_{\mathrm{max}} \equiv F_{\mathrm{av}}\frac{\mbox{observing
    timespan}}{N_{\mathrm{GP}} \times \mbox{ length of window} },
\end{equation}
\noindent
where $F_{\mathrm{av}}$ is the average pulsed $\gamma$-ray flux from the Crab
pulsar. $F_{\mathrm{max}}$ corresponds to the hypothesis that
\textit{all} $\gamma$-ray photons from the Crab pulsar come during GPs.  For our
choice of windows, $F_{\mathrm{max}}$ ranged from 60~$F_{\mathrm{av}}$ (for on-pulse
phase window) to 150~$F_{\mathrm{av}}$ (for IP window). 

The likelihood $p(N | F = F_{0})$ was calculated by running
simulations with different pulsed flux $F_0$ and computing the
fraction of runs with a number of matches $N$. The grid of trial flux values, in units of $F_{\mathrm{av}}$, was as follows: from 0 to 1 with the step of 0.25, from 1 to 20 with the step of 1 or 0.5, and then from 20 to 30 with the step of 5. For both choices of correlation window the probability density went down to 0 before $30\,F_{\mathrm{av}}$. Here we implicitly
assumed that a higher flux outside selected windows does not influence
the correlation within windows.

Since the number of simulation runs for each trial $F_0$ is finite, it
leads to an uncertainty in estimating the likelihood. We estimated
the statistical errors from the simulation using the following method.
Suppose that for some value of $F_0$ we have run $n$ simulations with
$y$ successes (i.e. cases where the number of matches in the
simulation equals the one obtained for real data, $N$). Then $y/n$
defines the estimate of probability of success $p$, which is also the
likelihood density $p(N | F = F_{0})$.  More precisely, $p|y$ has a
Beta distribution, 
with mean $(y+1)/(n+1)$ and variance
$\sigma_p^2=\frac{(y+1)(n-y+1)}{(n+3)(n+2)^2}$. We adopted $\sigma_p$
as an error of $p$ due to limited numbers of simulations performed.


However, there is another major source of uncertainty connected to
the fact that we record \textit{discrete} number of photons around
the GPs. Since photon detection is very well described as a Poisson
process \citep{ramanamurthy1998}, the error on detecting $N$ photons
in a certain window around GPs will be $\sqrt{N}$. In our case, for
all windows we had $N=1$, so to estimate the true value of the
likelihood we should take into consideration also the likelihood
curves for $N=0$ and $N=2$. These estimates for both windows are
plotted in Fig.~\ref{fig:correlation-bayes}, left. The shaded
region around each curve corresponds to $\pm \sigma_{p}$, calculated
by the above formula. For both correlation windows $N=1$ likelihood curves have maximum around 
$F_0/F_{\mathrm{av}}=1$, which means that, most probably, pulsed $\gamma$-ray flux
does not change during GPs (no correlation), or changes no more than few times (weak correlation).

With our limited data set we cannot say anything more about the exact value of $\gamma$-ray flux during GPs,
but we can place upper limits on it. On the grid of simulated fluxes $F_i$, one can convert the
continuous formula for posterior probability density (eq.~\ref{eq:bayes}) into a discrete
one for the probability that pulsed flux around GPs is \textit{less}
than $F_0$:
\begin{equation}
P(F \leq F_0 | N) = \frac{0.5\cdot \sum_{F_{i+1}\leq F_0}
  (p_i+p_{i+1})(F_{i+1}-F_i)}{0.5\cdot \sum
  (p_i+p_{i+1})(F_{i+1}-F_i)},
\end{equation}
where $p_i \equiv p(N | F = F_i)$.

To estimate errors in $P(F \leq F_0 | N)$, we assumed that our
uncertainty in $p_i$ due to a limited number of trials is much larger
than the error from calculating the integral as a sum. As one can see in 
Fig.~\ref{fig:correlation-bayes}, left, this simplification is reasonable.  Assuming all $\sigma_{p_i}$ are
independent, the uncertainty in $P(F \leq F_0 | N)$ is determined by
standard error propagation.

In Fig.~\ref{fig:correlation-bayes} we show the resulting
probabilities that the $\gamma$-ray flux from the Crab pulsar during
GPs does not exceed a given number of times the mean flux reported by
\cite{FermiCrab2010}. Errors due to the limited number of simulation
runs are plotted as errorbars, whereas those due to a discrete number
of GP/photon matches are given by separate posterior probability
curves for $N=0$ and $N=2$.

Obviously, the smaller the correlation window for a fixed observation
timespan and the smaller the number of GPs in the sample, the larger
$F_{\mathrm{max}}$, and the broader is the resulting posterior
probability density. That is why for our data set we could obtain the
posterior probability densities only for correlation windows which
included the IP, because most of GPs come within this phase range. For
the main pulse GPs, $p(F=F_0|N)$ is very broad, having almost the same
probability density up to $\sim 100 F_{\mathrm{av}}$. For the same
reasons, the analysis on the sample of GPs corrected for refractive
scintillation did not give any meaningful results.

\section{Conclusions}\label{conclusions}

\textit{No obvious correlation was found between \textit{Fermi} photons of
  energies $>$ 100~MeV and radio giant pulses at the frequency of
  8.9\,GHz.}  No change in the Crab GP generation rate was found on
timescales from 10 to 120\,s around $\gamma$-ray photons and with any
possible lag within $\pm 40$\,min with respect to $\gamma$-ray photons.

With 95\% probability, the high energy flux of the Crab pulsar during
GPs is less than 4 times the average $\gamma$-ray pulsed flux for the
on-pulse (MP+IP+bridge between them) phase window. For IP GPs only,
the 95\% upper limit on $\gamma$-ray flux in the IP phase window is 12
times the average pulsed flux.  If we consider the uncertainty due to
discrete numbers of matches between photons and GPs, the 95\% upper
limits are 3$-$5.5 times the average pulsed flux for the pulsed
emission window, and 8$-$16 for the IP window.

A few explanations may be offered for the non-detection of GP-$\gamma$-photon correlations. The most natural is that production of GPs
depends on non-stationary changing coherence conditions, which vary by
a large degree even for similar magnetospheric particle densities.
Another possibility is that beaming in radio and at high energies are
somewhat different, so that simultaneous GPs and $\gamma$-ray photons are emitted in
different directions. 

Overall, our results suggest that enhanced pair
creation is not a dominant factor for GP occurrence, at least for high
frequency IP GPs. However, our flux increase estimations are not on
the level of a few percent, as in the work of \cite{shearer2003} at
optical wavelengths. To reach that sensitivity we need more data,
which will help push down the upper limit on flux during GPs and will
make possible the analysis on subsamples of GPs, such as the brightest
ones. Also, including radio frequencies below 4\,GHz is potentially
interesting not only for investigating the correlation for MP GPs
separately (MP GPs are much more common at lower frequencies), but
also for re-doing the analysis for low-frequency IP GPs, since they
might be generated by different physical processes than the
high-frequency IP GPs \citep{moffett1997}. All these questions are
being investigated with our ongoing radio observation campaign using
the 42-ft telescope at the Jodrell Bank Observatory (UK) and the
140-ft telescope at the Green Bank Observatory (WV).

\bigskip 

\begin{acknowledgments}
  The National Radio Astronomy Observatory is a facility of the
  National Science Foundation operated under cooperative agreement by
  Associated Universities, Inc. This work was supported by \textit{Fermi} grant
  NNX10AD14G.  M.~A.~McLaughlin is an Adjunct Astronomer at the National Radio Astronomy Observatory and is supported by a Cottrell Fellowship, a Sloan Fellowship, and a WVEPSCOR Research Challenge Grant. 
  A.~V.~Bilous thanks T.~T.~Pennucci (UVa) for the useful comments on the paper draft. 
\end{acknowledgments}

\bibliographystyle{apj}
\bibliography{paper_bibliography}

\end{document}